*Research Article*

# Integration of Blockchain and IoT: An Enhanced Security Perspective

**Mahdi H. Miaz[1,2],\* and Maaruf Ali[3]**

[1]Xiamen University Malaysia, Selangor, Malaysia
m.miraz@ieee.org
[2]Wrexham Glyndŵr University, Wrexham, UK
m.miraz@ieee.org
[3]Epoka University, Tiranë, Republika e Shqipërisë (Albania)
maaruf@ieee.org
\*Correspondence: m.miraz@ieee.org

**Abstract:** Blockchain (BC), a by-product of Bitcoin cryptocurrency, has gained immense and wide scale popularity for its applicability in various diverse domains – especially in multifaceted non-monetary systems. By adopting cryptographic techniques such as hashing and asymmetric encryption - along with distributed consensus approach, a Blockchain based distributed ledger not only becomes highly secure but also immutable and thus eliminates the need for any third-party intermediators. On the contrary, innumerable IoT (Internet of Things) devices are increasingly being added to the network. This phenomenon poses higher risk in terms of security and privacy. It is thus extremely important to address the security aspects of the growing IoT ecosystem. This paper explores the applicability of BC for ensuring enhanced security and privacy in the IoT ecosystem. Recent research articles and projects/applications were surveyed to assess the implementation of BC for IoT Security and identify associated challenges and propose solutions for BC enabled enhanced security for the IoT ecosystem.

## 1. Introduction

The primary aims of this article is to conduct a detailed study exploring how (and to what extent) Blockchain technologies can be utilised in enhancing the overall security of the Internet of Things (IoT) ecosystems and to foresee the future of Blockchain of Things (BCoT) fusion. Considering the fact that the Blockchain is comparatively an avant-garde technology, this paper presents a representative sample of research conducted in the last ten years, commencing with the early work in this domain. Although, identifying how the Blockchain can further enhance the security paradigm of IoT is the main focus of the paper, to do so various other usages of the Blockchain and similar digital ledger technologies were explored along with their applications, impediments, privacy and security concerns. This paper explicates our previous work [1] presented at the International Conference on Emerging Technologies in Computing, which was held at London Metropolitan University in 2018.

Like many other domains of computing, security and privacy issues are the major concerns of the Internet of Things (IoT) ecosystem [2, 3]. To fortify the backbone for improved security and privacy of IoT, the Blockchain is considered to be able to play a vital rôle. In fact, Blockchain research has become truly multifaceted as researchers from both industry as well as academia are applying the Blockchain in new dimensions on a regular basis. In the Proof-of-Work (PoW) concept [4], an algorithm based on mainly solving a mathematical challenge, is the major method to assure the security aspects of the BC by recording and maintaining a complete digital ledger of all the completed transactions. These transactions are thus unalterable. A high-level system block diagram of how the BC technology works is shown in Fig. 1.

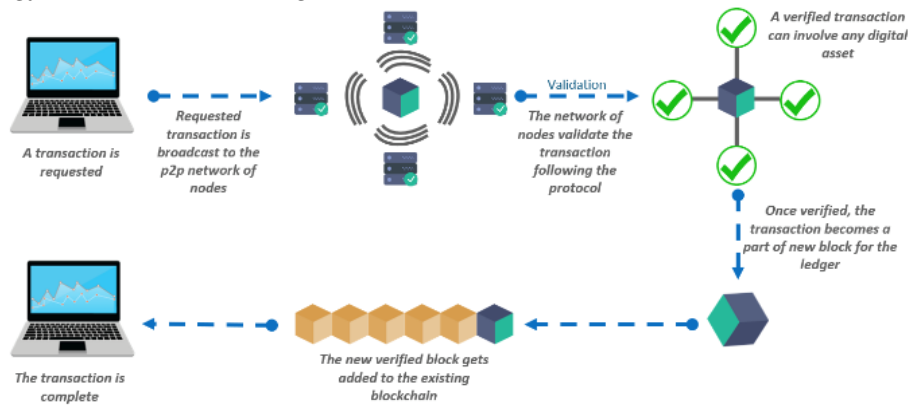

**Figure 1.** A high-level view of the Blockchain Technology [5].

In addition to this, the BC also takes advantage of the Public Key [5], which is purposely made chaotic in nature for ensuring the highest level of security, in order to register the identity of the users. Thus, an extra layer of privacy is ensured automatically. As evident by many research and project reports, the adoption of the Blockchain technology has been found to be successful in many non-monetary domains such as in the supply chain, healthcare systems, online/electronic voting, proof of location, distributed cloud storage, securities settlement [6, 7], even in human resource management and recruitment [8]. Furthermore, implementation of a blockchain may be seen as a six-layer hierarchical model [9], as shown in Fig. 2, below.

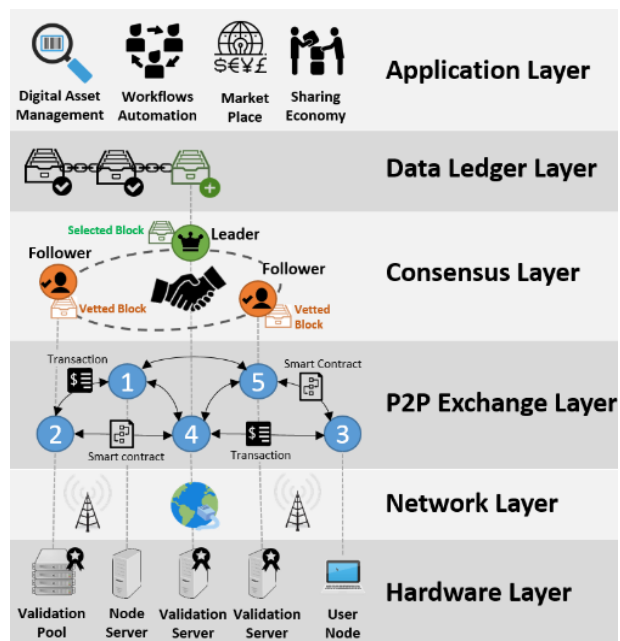

**Figure 2.** The Layered Model of the Blockchain Technology [9]

The authors of this paper not only surveyed research articles but also considered relevant projects/applications to ascertain the applicability of Blockchain technology for augmented IoT security and to distinguish the challenges associated with such application of the BC and thence to put forward probable solutions for BC enabled enhanced IoT security systems.

The knowledge domain of the research is in the realm of the Internet of Things (IoT), Internet of Everything (IoE), Wireless Sensor Network (WSN) and Distributed Digital Ledger (DLT), specifically, in Blockchain and crypto-currency.

## 2. Blockchain Fundamentals

To understand how Blockchain can be applied for enhancing IoT security, it is very important to understand how these two technologies are put into function. In this section, the basic technological fundamentals of Blockchain have been briefly described while the next session introduces the Internet of Things (IoT) ecosystem.

A Blockchain mainly consists of two separates but interrelated integrant. These are as follows:
1. **Transaction:** in a digital ledger system such as the Blockchain, a transaction is basically the action triggered by the participant.
2. **Block:** A block, in a Blockchain system, is a collection or pool of data which records the transaction and other relevant details such as the correct sequence, timestamp of creation, *et cetera*.

Based on the scope of how a Blockchain is going to be used, it can be of two types: private or public. In a public Blockchain, usually all the users have both read and write permissions. One example of such a public Blockchain application is in recording the generation and financial flow of the Bitcoin cryptocurrency. However, there are also some public Blockchains where access is limited to either write or read rights, depending on the rôle of the user in the system. The aim of a private Blockchain, on the contrary, is to conceal the details of the users. To ensure that, access is limited to some trusted participants or members of a single organization. A Blockchain that is controlled by a consortium is known as a "consortium blockchain" [9]. This is particularly pertinent amongst governmental institutions and allied sister concerns or their subsidiaries thereof.

The implementation of the Blockchain technology being mainly public puts the BC at the forefront of other technologies, especially in terms of security and transparency aspects. Since each participating node possesses its own copy of the complete blockchain i.e. whole blocks of updated records and transactions, the data thus remain unaltered. Any unauthorised or unexpected changes will thus be publicly verifiable. However, the data recorded in such publicly available blocks are hashed and encrypted (by the private key) to ensure security and anonymity. Because the private key is used to encrypt the data, it cannot be publicly scrutinised nor interpreted.

Although a centralised implementation of the Blockchain technology is possible, it is mostly decentralised in nature, which is considered to be another one of its major advantages. It is decentralised in the sense that:

- The data, comprising the transactions and associated blocks, are distributed among the participating nodes of the Blockchain network, rather than storing them in a single piece of node or storage device.
- The transactions are approved by a set of specific rules or algorithms, thus eliminating the influence of being biased by one single authority involving substantial trust in order to reach a consensus.
- The Blockchain systems only allows new verified blocks to be appended to the old chain. As the previously added blocks are already public and distributed, they are openly verifiable and hence cannot be altered or revised. Thus, the overall security of a Blockchain ecosystem is another advantage over other technologies.

Once a transaction is triggered by a participant, it is not added straightaway to the chain of blocks i.e. the blockchain. In order for a newly initiated transaction to be appended with the existing chain, the transaction has to go through the validation and verification processes. The participating nodes of the Blockchain networks must apply a set of predefined rules or specific algorithms for this purpose. The set of rules or algorithms basically defines what is perceived as "valid" by the respective Blockchain system and may vary from one to another. Rather than adding one single transaction in a block, usually a number of such transactions are put together in order to construct a new block. This newly prepared block is then sent to all other participating nodes of the Blockchain network so that they can be appended to their copy of the existing chain of blocks. Each succeeding block of the chain comprises a hash, a unique digital fingerprint, of the preceding one.

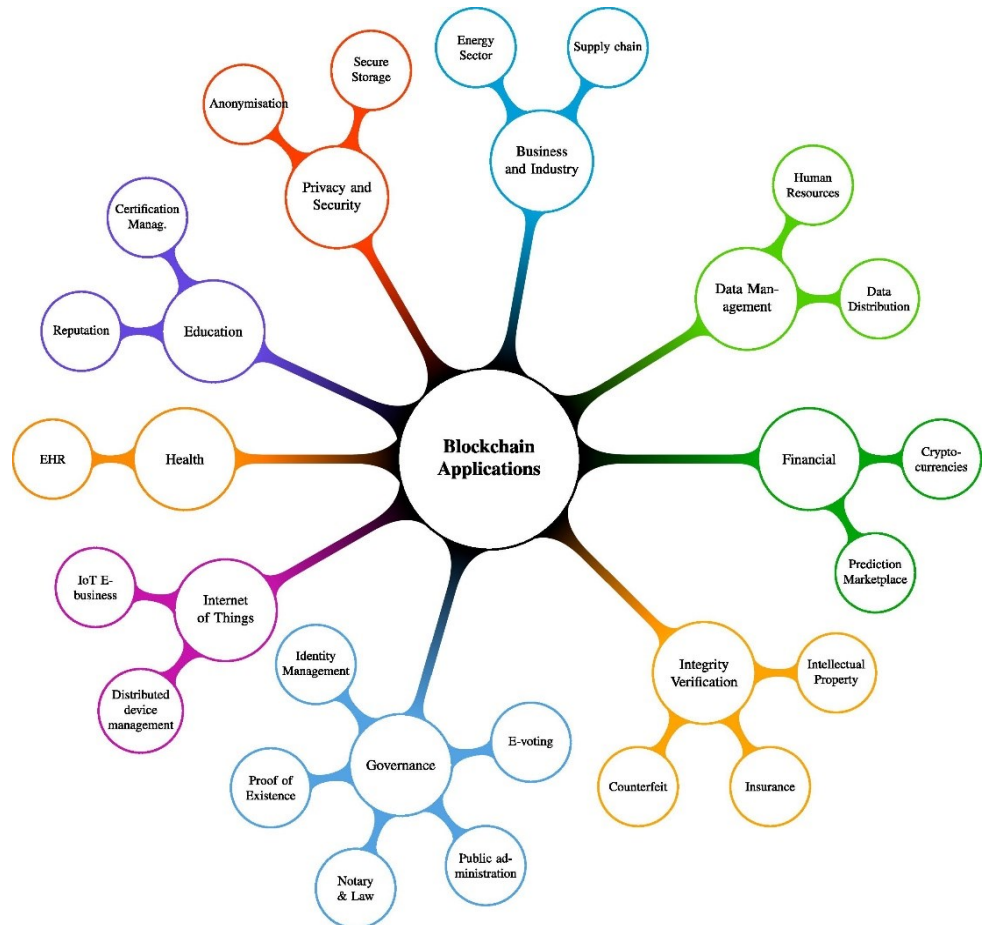

**Figure 3.** Typical Application of the Blockchain Technology [10]

The Blockchain not only verifies and validates all the newly triggered transactions but also maintains an irreversible lifelong record of them, while assuring that all the identification related information of the users or the participants are kept incognito. All the personal information of the users is sequestered while substantiating all the transactions. This is achieved by reconciling mass collaboration by cumulating all the transactions in a computer code based digital ledger. Thus, in a Blockchain system, instead of trusting each other or an intermediary, the participants need to trust the decentralized network system itself. The Blockchain itself then has become the ideal "Trust Machine" [11, 12] paradigm.

Although the Bitcoin cryptocurrency first used the Blockchain, it is considered to be just an exemplary use of the BC. Blockchain technology is a relatively novel technology in the domain of computing that is enabling illimitable applications, such as in and not just limited to: healthcare systems, human resource management, recruitment, storing and verifying legal documents including deeds and various certificates, IoT and the Cloud. In fact, Tapscott [13] has perfectly connoted Blockchain to be the "World Wide Ledger", facilitating many novel applications beyond just the simple verifying of transactions such as in: recording smart deeds, decentralized and/or autonomous organizations/government services *et cetera*. Fig. 3, shows the typical and diverse applications of the blockchain technology.

The level of pervasiveness of the application of blockchain technology can be seen illustrated by Fig. 4, which shows the diverse application just in the biomedical domain. Fig. 4 also lists six reasons for using the blockchain: access control, non-repudiation, data versioning, data integrity, data auditing and for data provenance.

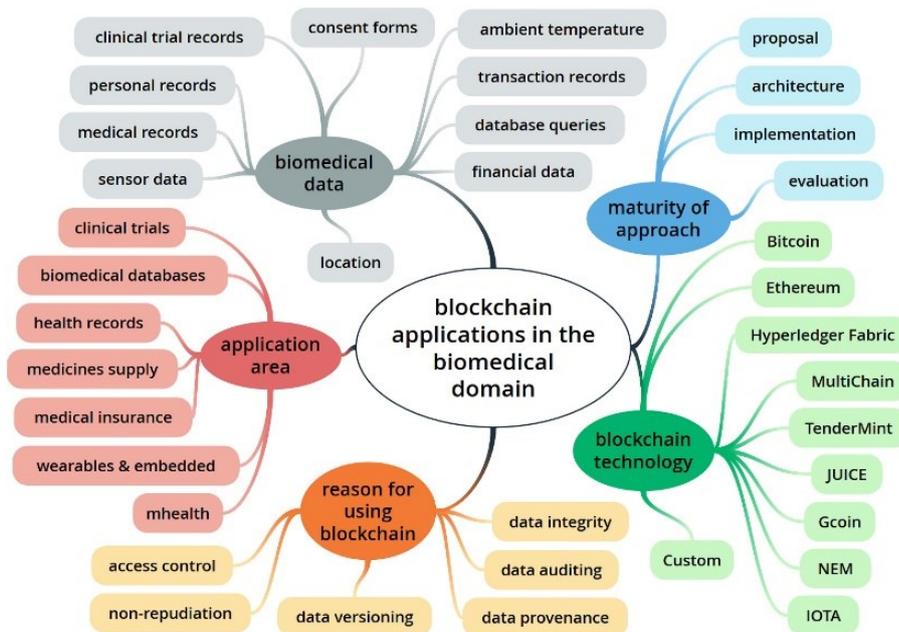

**Figure 4.** Mindmap of the Blockchain Applications in the Biomedical Domain [14].

## 3. Internet of Things (IoT)

The term 'Internet of Objects' or 'Internet of Things' (more commonly referred to as 'IoT') - denotes the electronic or electrical devices of many different sizes and capabilities connected to the Internet. This connection is mainly by using wireless sensors, but excluding those primarily involved in communications with human beings, i.e. the traditional Internet. New IoT devices are being marketed on a regular basis and thus the scope of the connections is ever broadening beyond just basic machine-to-machine communication (M2M) [15].

There are many types of IoT devices employing a wide range of applications, protocols, and network domains [16]. The growing preponderance of IoT technology is enabled by the physical objects being connected to the Internet by various types of short-range wireless technologies such as sensor networks, RFID, ZigBee and through location-based technologies [17].

The emergence of IoT as a distinctive entity was reached (according to the Internet Business Solutions Group (IBSG)) when more inanimate objects were directly connected to the Internet bypassing human users [18]. This accelerating process has been gaining momentum ever since the rollout of CISCO's 'Planetary Skin', the Smart Grid and intelligent vehicles [18]. IoT is already on the verge of making the Internet truly pervasive, with devices already embedded into consumer white goods, including personal and intimate devices [19] in our daily lives. IoT devices are only standardised in their use of the Internet networking protocols and not how they interface to the Internet or with each other. This immediate potential inhibiting factor needs to be addressed.

IoT may be deployed with added privacy, security and management features to link, for example, vehicle electronics, home environmental management systems, telephone networks and control of domestic utility services. The broadening scope of IoT and how it can link with heterogeneous networks is shown in [18].

A standard IoT ecosystem typically comprises of the following five components:
1. **Sensors**: sensors are mainly responsible for the collection and transduction of the required data;
2. **Computing Node**: such nodes containing the central processing unit (CPU), are required for processing the data and information received from the sensor(s);
3. **Receiver**: which is actually a transceiver, facilitates the collection of the message sent by the local and remote computing nodes or other associated devices;
4. **Actuator**: which could be electro-mechanical in nature, works on the basis of the decision taken by the Computing Node, processing the information received from the

   sensor and/or from the Internet, then triggering the associated device to perform a function; and
5. **Device**: to perform the desired task as and when triggered.

## 4. BC Enabled Enhanced IoT Security

In an IoT ecosystem [20, 21], most of the communication is in the form of Machine-to-Machine (M2M) interactions, that is, without any human intervention whatsoever. This means that the establishment of trust amongst the participating machines is a big challenge that IoT technology still has not met extensively. However, the application of the Blockchain may act as a catalyst in this regard, by enabling enhanced scalability, security, reliability and privacy [11, 12]. This can be achieved by deploying Blockchain technology to track billions of devices connected to the IoT ecosystems and then used to enable and/or coordinate transaction processing. In fact, a specific search engine already exists, called "Shodan", that describes itself as "the world's first search engine for Internet-connected devices"[1]. The use of this search engine by anyone will also expose any insecure IoT devices and hence their need for rectification. Application of the Blockchain in any IoT ecosystem will further enhance the reliability by completely eliminating any Single Point of Failure (SPF). In Blockchain, data is encrypted using cryptographic algorithms as well as the hashing techniques. Thus, the application of Blockchain in an IoT ecosystem can offer better security services. However, to perform the hashing techniques and implement the cryptographic algorithms, the systems shall obviously demand more processing power, which IoT devices currently lack. Further research is required to overcome this present limitation, including extending the longevity of the in situ powering source.

Underwood [22] considers the application of Blockchain technology to completely overhaul the digital economy. Ensuring and maintaining trust is both the primary and initial concern of the application of the blockchain. BC can also be used to gather chronological and sequence information of transactions, as it may be seen as an enormous networked time-stamping system. For example, NASDAQ is using its 'Linq blockchain' to record its private securities transactions. Meanwhile the Depository Trust & Clearing Corporation (DTCC, USA) is working with Axoni in implementing financial settlement services such as post-trade matters and swaps. Regulators are also interested in BC's ability to offer secure, private, traceable real-time monitoring of transactions.

Securing operational technology is also of paramount importance. Blockchain can help to prevent tampering and spoofing of data by managing and securing industrial IoT and operational technology (OT) devices. Once a sensor, device or controller has been deployed and is working, it cannot be tampered with - since any compromised devices will be recorded in the BC.

Since IoT highly utilises the existing wireless sensor network (WSN) technologies, intrinsically it remains vulnerable to privacy as well as security threats. On the contrary, blockchain, by its design and architecture-consensus method and cryptographic techniques – is considered as a Trust Machine [23, 24]. Thus, it possesses the potentials to address major share of the security issues found in IoT. Miraz [24] argues them to be complementary technologies for each other: BC requires participating nodes for consensus approach which can be supplemented by IoT devices while IoT requires security features which can be met by BC such as: transparency, privacy, immutability, operational resilience and so forth.

IoT is a cyber-physical system which help represent the "connected" physical world into part of a substantial realm of information system – the cyber world. However, due to various reasons, the security aspects of IoT has not been properly addressed at the design phase of the devices and products. With the advent and increasing popularity of BC, there has been a paradigm shift in IoT research, particularly integrating IoT and BC [25, 26] together for a more robust but secure cyber world. However, since the technologies are still not fully mature, many challenges are yet to be addressed arising from such an integration [27]. Many studies [28, 29, 24] suggest applications of BC as a probable solution to tightening the security aspects of the IoT ecosystem including the protection from the "Stalker" [28] attack.

---

[1] https://www.shodan.io/

Since IoT is built on the foundation laid by the wireless sensor network (WSN) [30], characteristically each node of an IoT ecosystem is considered to be prone to attacks such as from the Distributed Denial-of-Service (DDoS) [31, 32] threat and if compromised, it may serve as a point of failure. IoT networks are mostly leveraged on the cloud environment. Such centralised architecture suffers from a Single Point of Failure (SPF) and further adds to vulnerability.

IoT devices gather and/or generate a vast amount of data which are communicated over the Internet for processing and in decision making purposes. Data privacy and authentication is considered to be a constant critical threat for the IoT environment. In the absence of proper security measures, these vast amounts of data can be mishandled and used inappropriately [33]. It is thus extremely important to safeguard the IoT system from injection attacks and spoofing. As the name implies, an injection attacks tries to inject false data or measures into the system and thus affect the overall decision-making process.

Within the recently materialised Machine Economy, the data generating sensors are now able to efficiently share, exchange and merchandise the data in various autonomous systems and marketplaces. However, boosting trust amongst the partaking entities remains a core challenge. Existence of a publicly verifiable system for audit-trail, without the need for a third party, for establishing trust, is prudent to address the non-repudiation problem [34]. Thus far, several such blockchain based applications have been piloted: FileCoins [35] and TransActive Grids [36] are two major examples. Utilising these applications, the devices can generate money by trading.

[37] adopts pseudonymization approach which slices the data into several small parcels before they are sent to the smart devices of an IoT enabled ecosystem such as smart home. The BC records the hashes of the data generated by the smart IoT devices and act as a certificate provider. Only the owner can rebuild the data and assign access rules of the data for the smart devices or external entities (service providers) based on public keys. The framework also integrates the BC into the different layers, as shown in Fig. 2, (application, database, communication and physical layers) and thus addresses the various security related limitations of these layers. The Ethereum platform has been utilised to provide smart contract functionalities applying BC as a distributed Database. The BC enabled application layer provides an extra layer of security – by denying intruders any illegitimate access to the dependent processes.

The Filament crypto platform developed an alternative solution by the application of the Telehash[2] protocol to adopt BC [38]. The prototype developed by Filament, by utilising BC as well as smart contracts, enables smart IoT devices to discover, interact and communicate (message exchange) with each other with the freedom to act independently without the need for any central control. Prior to any communication taking place, the participating devices are required to satisfactorily authenticate themselves to the other parties by utilising the appropriate security protocols, such as Transport Layer Security (TLS) and Secure Socket Layer (SSL), through public key infrastructure (PKI). In contrast to the models developed by [37] and [38], Prabhu et al. [39] adopted a different approach. Applying BC as the backbone of the IoT ecosystem, IP addresses were rather utilised to retrieve data recorded on the distributed ledger. Their system also utilises the events stored on the ledger for notifications.

In an IoT environment, some of the intermediate devices may act as hops. Therefore, designing a secured architecture realising the importance of the private communication protocols is a dire necessity. For secure communications amongst the smart devices of an IoT ecosystem, the Moeco [40] prototype has been developed by a Moeco – a Berlin, German based start-up. Moeco adopts the novel notion of "Domain Name System (DNS) of things", by utilising BC for the purpose of data routing in IoT environment. While the current version of Moeco is based on the Ethereum platform, future research plan involves adopting Exonum [41] – a custom-built Byzantine consensus approach.

Hashemi et al. [42] introduce a publish–subscribe approach-based solution to IoT data security, analogous to what [37] aimed to achieve. The objectives of [42] are twofold:
1) To segregate data store from data management; and
2) To architect the components adopting a distributed decentralised but scalable approach.

---

[2] https://github.com/telehash

Amongst the three layers modelled in the prototype [42], BC has been utilised at the Data Storage system for enhanced transparency and persistent distribution. Further application of BC is implemented at the Data Management procedure where access-controlled data is collected in a decentralised environment via BC enabled rôle-based interactions such as data source, owner, endorser and requester.

Thus far, one of the still major challenges of securing IoT data is implementing an authentication mechanism and access control which perfectly meets the special requirement of the decentralised nature of the IoT ecosystem comprising low processing powered devices. Classical Access Control Protocols (ACLs) as well as others including Discretionary Access Control (DAC), Mandatory Access Control (MAC) and Attribute-Based Access Control (ABAC) do not fit well in the IoT environment with its decentralised architecture. To address this problem, a BC and smart contract based solution has been proposed by Deters [43]: data or resource owner, signing using appropriate credentials, sends an "Announcement" transaction to the BC ecosystem for recording on the ledger, while the recorded data can only be accessed through sending a request transaction to a smart contract that oversees the access control.

Bahga and Madisetti [44] presented a new approach to implement the Blockchain Platform for Industrial Internet of Things (BPIIoT) for Cloud-based Manufacturing (CBM) systems that aims to provide manufacturing resources and capabilities as a cloud service. This Ethereum based system also utilised smart contracts to facilitate its services, whilst enhancing the security of the transactions through changeable public keys. Korpela et al. [45] examines the supply chain specific functionalities and requirements for integration of cloud computing, IoT and BC. While integration of cloud computing with IoT can enhance the operation of interoperable digital supply chains by enabling cost-effective business models, BC can play a vital rôle in achieving disruptive digital transformation in supply chain networks.

BC enabled IoT systems can also be implemented in the smart energy exchange aspect of Industry 4.0 [8], especially in M2M communication for which it is particularly designed for this purpose [46]. The proposed system [46] enables the participating smart devices to trade any type of commodity such as steam, natural gas etc. using BC. An energy producer publishes the price using BC transactions while the consumer purchases the best-published offer using the relevant crypto asset.

Dorri et al. [47, 48] proposed a BC enabled multitier architecture with enhanced security and privacy, satisfying the IoT specific requirement while eliminating BC's consensus specific limitations. Since BC's PoW consensus approach is computationally expensive and highly critiqued for its stand against green computing notion [49], rather an access policy-based approach managed by a single miner is adopted – an overlay network as well as cloud storage have been utilised. When a new node (smart device in this case) is added to the network, a miner generates a new block corresponding to the newly added node which contains two distinguished headers: one to link to the previous block in the chain and the other to contain the data access policy. Unlike other BC applications, to enhance privacy and confidentiality, symmetric keys (Diffie–Hellman algorithm) are used to communicate amongst the nodes, maintained and distributed by the miner. Thus, the single miner becomes a powerful central authority and the decentralised aspect of BC is compromised to some extent.

Securing the IoT environment is not the only reason for integrating with BC, the fusion also provides the benefits of smart contracts, enabling innovation of new business models as well as solving existing problems. However, the breaches brought by the vulnerabilities associated with smart contracts [49] are also need to be considered. The system proposed by Wörner et al. [50] enables sensors to trade data for Bitcoin. Addresses of each participating sensor nodes are Bitcoin's public keys. To request (purchase) data from a sensor, a transaction (including transfer of Bitcoin) to the sensor is announced. Upon verification, the sensor will then create a response transaction for the requesting node, with the data requested. In fact, this [50] is a generalisation of the IoT electric business model proposed by Zhang and Wen [51].

The Enigma framework, proposed by Zyskind et al. [52], is a P2P that enables different parties to both store and operate on the data while keeping it completely private. The purpose is to securely distribute data amongst various nodes and segregate the data from their references, thus making it difficult to rebuild the original data. This is achieved through the administration of the data storage using a Distributed Hash Table (DHT) for the shared secret data chunks as well as implementation of external BC to govern access control, network monitoring and identities. A similar approach, BC enabled auditable storage system for IoT Data, has been proposed by Shafagh et al. [53].

IOTA cryptocurrency, developed by the IOTA foundation, is an open-source micro-payments platform paying particular attention to IoT. However, instead of using traditional BC, IOTA implements TANGLE - a BC without having any blocks or even miners [54], rather leveraging on acyclic graphs, making it highly scalable.

A two-layered (level 0 and level N) BC based IoT security framework has been designed by Chakraborty et al. [55] taking into consideration the resource-constrained nodes. Devices, with limited processing capabilities to enforce security primitives, belongs to level 0. On the contrary, level N comprises two types of nodes: primary and secondary. While the primary nodes administer the processing of data, communication, access control and other associated tasks, the rôle of secondary nodes is to assist the primary ones. Due to the resource-constrained feature of the level 0 nodes, they are barred from directly communicating with other level 0 nodes. However, indirect communication via other level N nodes, capable of enforcing security measures, is in place.

## 5. Concluding Discussion

To answer the research question "To what extent can the Blockchain be used in enhancing the overall security of the Internet of Things (IoT) ecosystems?", this paper first introduced how these two emerging technologies work. The current security issues related to IoT systems were also discussed. The authors of the article then investigated how the application of the Blockchain can eliminate these security concerns inherent in the IoT ecosystem and improve its overall security.

Both the Blockchain and Internet of Things (IoT) are two relatively new but promising technologies being successfully used in multifaceted applications. The way the application of the Blockchain has widened beyond its initial use for Bitcoin generation and dealing has conclusively shown its relevance and versatility in general networked secure transactions. IoT also proved itself to be capable of doing far more things than being a simple wireless sensor network. In fact, Blockchain and its variants combined offers many security aspects such as enhanced privacy, stronger security, full traceability, inherent detailed data provenance and accurate time-stamping which other technologies still could not offer as standalone features. BC has seen its adoption beyond its initial application areas and is now used to secure any type of transaction, whether: human-to-human (H2H), machine-to-machine (M2M) or human-to-machine (H2M) communications. The adoption of Blockchain appears to be secure, especially allied with the world emergence of the Internet-of-Things (IoT). Its decentralized application across the already established global Internet is also very appealing, in terms of ensuring network redundancy, data redundancy through distribution and hence survivability.